\documentclass[conference,compsoc]{IEEEtran}

\usepackage{graphicx}
\graphicspath{{figure/}}
\usepackage{xcolor}

\begin{document}
%
\title{Trust and Reputation Management for Blockchain-enabled IoT}

\author{
    \IEEEauthorblockN{
    Guntur Dharma Putra\IEEEauthorrefmark{1}\IEEEauthorrefmark{4},
    Sidra Malik\IEEEauthorrefmark{1}\IEEEauthorrefmark{2},
    Volkan Dedeoglu\IEEEauthorrefmark{2},
    Salil S. Kanhere\IEEEauthorrefmark{1}\IEEEauthorrefmark{4}
    and Raja Jurdak\IEEEauthorrefmark{3}
}
\IEEEauthorblockA{
    \IEEEauthorrefmark{1}UNSW Sydney
    \IEEEauthorrefmark{2}CSIRO's Data61
    \IEEEauthorrefmark{3}QUT, Brisbane
    \IEEEauthorrefmark{4}CSCRC, Australia \\
    \{gdputra, salil.kanhere\}@unsw.edu.au, \{volkan.dedeoglu, sidra.malik\}@data61.csiro.au, r.jurdak@qut.edu.au
}
}


\maketitle

\begin{abstract}
In recent years, there has been an increasing interest in incorporating blockchain for the Internet of Things (IoT) to address the inherent issues of IoT, such as single point of failure and data silos. However, blockchain alone cannot ascertain the authenticity and veracity of the data coming from IoT devices. The append-only nature of blockchain exacerbates this issue, as it would not be possible to alter the data once recorded on-chain. Trust and Reputation Management (TRM) is an effective approach to overcome the aforementioned trust issues. However, designing TRM frameworks for blockchain-enabled IoT applications is a non-trivial task, as each application has its unique trust challenges with their unique features and requirements. In this paper, we present our experiences in designing TRM framework for various blockchain-enabled IoT applications to provide insights and highlight open research challenges for future opportunities.
\end{abstract}


\IEEEpeerreviewmaketitle

\begin{figure*}
    \centering
    \includegraphics[width=\textwidth]{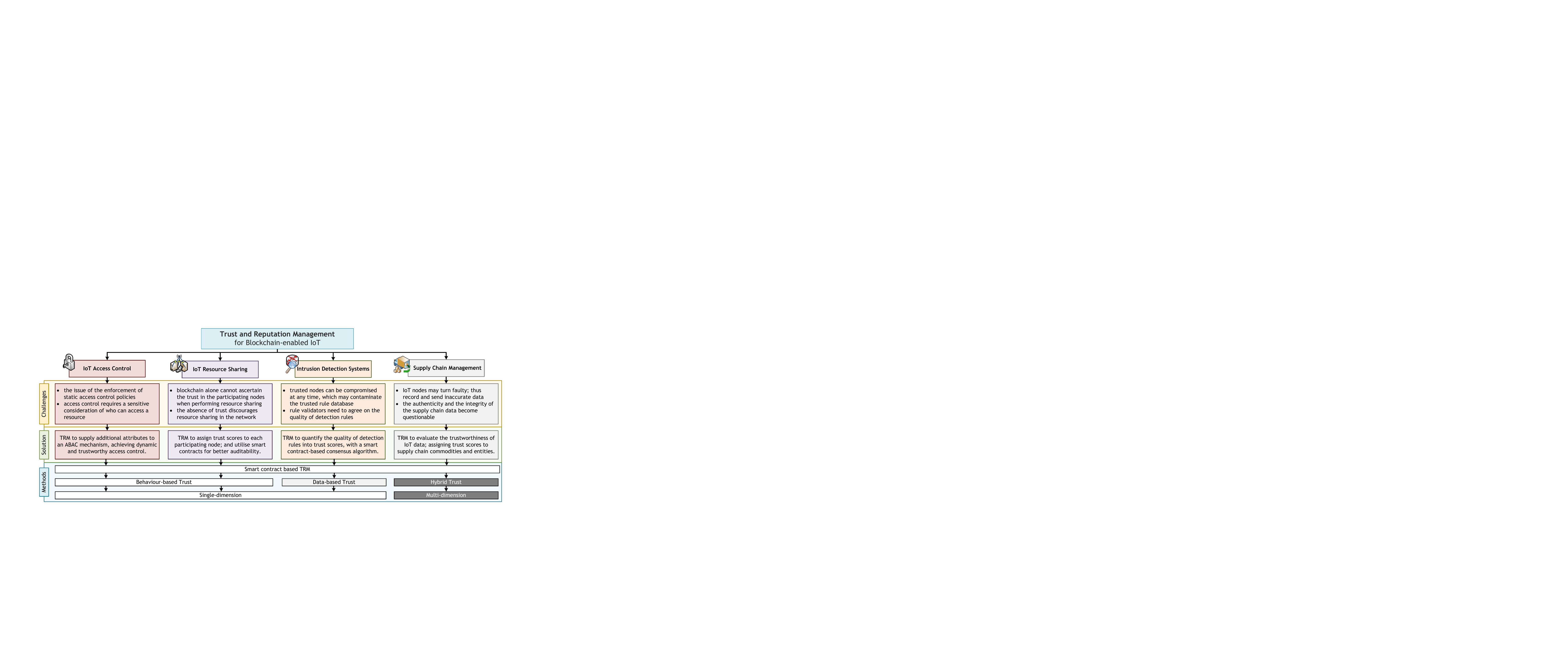}
    \caption{The summary of our research experiences in designing TRM framework for various IoT applications. While there is convergence in the employed TRM techniques and methods, each application has its own trust challenges with their unique features and requirements, which significantly influence the final TRM framework. We explain our experience in more detail in Section~\ref{sec:applications}.}
    \label{fig:journey-summary}
\end{figure*}

\section{Introduction}
Blockchain, the underpinning technology of cryptocurrency, has shown its potential to revolutionise the Internet of Things (IoT)~\cite{dai2019b}. The salient features of blockchain, such as decentralisation and auditability, have solved the inherent challenges of IoT, such as centralisation, data silos and interoperability~\cite{dedeoglu2020a}. For example, in the context of supply chain management, blockchain integration with IoT has increased the transparency of how the traded commodities are handled from production to the retail shelves, thus increasing traceability~\cite{malik2018}.

Although distributed consensus algorithms remove the need to trust a centralised entity, blockchain by itself cannot ascertain the veracity of the collected observation data~\cite{dedeoglu2019}, thus raising concerns about the integrity and authenticity of the logged data. Moreover, blockchain mechanisms cannot guarantee the trustworthiness of data at the origin, as the data is collected from the physical world by an entity that may be malicious or prone to errors~\cite{dedeoglu2019}. In fact, invalid observational data cannot be corrected once stored on-chain, which would overload the blockchain with immutable bad data.

Blockchain-enabled IoT networks utilise authentication and authorisation schemes to limit access only to only authorised nodes. However, these schemes can not ascertain in situ reliability and trustworthiness of authorised nodes, as these schemes do not monitor the behaviour of nodes over the operational period. The case of Mirai Botnet exemplifies how IoT nodes can be compromised post-authentication and become malicious~\cite{antonakakis2017}, which could severely impede the security and resiliency of the network. As critical infrastructures are being increasingly connected to IoT networks~\cite{ahmad2020}, the presence of malicious adversaries could eventually cause severe detrimental effects, as exemplified in the recent Colonial Pipeline attack, where a ransomware attack halted a major gas pipeline in the US~\cite{colonial-attack2022}.

Trust and Reputation Management (TRM) is an effective approach to overcome the aforementioned trust issues, wherein the trustworthiness of each node is continuously evaluated. TRM performs trust evaluation by processing feedback and ratings from other nodes, after which TRM quantifies the nodes' trustworthiness~\cite{hasan2022}. In TRM, each node is assigned trust or reputation scores that represent its trustworthiness level, using which other nodes in the network may conveniently infer their  trustworthiness level and the data they generate. TRM can benefit from the incorporation of blockchain with its salient features, such as transparency, tamper-resilience, verifiability, and pseudonymity. These features improve TRM resiliency and address the aforementioned issues of conventional TRM~\cite{Bellini2020}. For instance, the decentralisation inherent in blockchain eliminates the reliance on a Trusted Third Party. Trust related data can also be stored on the shared immutable ledger, maintaining integrity and high availability. Smart contracts can be employed to enforce trustworthy collection of collaboration evidence and trust calculation.

However, designing a blockchain-based TRM to provide strong assurance of trustworthiness in data and participants behaviour is a non-trivial task. There exist many challenges in incorporating TRM for blockchain-enabled IoT, which specifically depend on the IoT applications, as each application has its trust challenges with their unique features and requirements.

In this paper, we present our research experiences in designing TRM framework for various applications of blockchain-enabled IoT, namely access control, intrusion detection systems, supply chain management and network resource sharing. First, we describe the definition of trust and reputation with its subtle differences and outline the fundamental stages in TRM. Subsequently, we discuss the salient features of blockchain that would improve the design of TRM. We highlight specific trust issues in each application domain that necessitates the incorporation of TRM for addressing these specific challenges. We summarise the lessons we learned in the process, which can guide future research in this area. Lastly, we highlight several open research challenges in blockchain-based TRM for future work. We present a summary of our research experiences in Fig.~\ref{fig:journey-summary}.

\section{Overview of Trust and Reputation Management}
\label{sec:trm-overview}
Trust represents a subjective belief towards an entity (trustee) regarding its behaviour, which is gradually built from repetitive interaction~\cite{Batwa2021}. The higher the trust to a trustee, the more it is expected to do things that would benefit the trustor, i.e., an entity that trusts the trustee. On the other hand, reputation can be seen as a collection of trust opinions towards an entity, which represents a global view of an entity's trustworthiness~\cite{putra2021d}.

TRM is designed to quantitatively evaluate the accountability of each node in an IoT network by assigning numerical scores to each node which represents its trustworthiness level~\cite{Bellini2020}. TRM works by collecting direct experience or feedback from the peer nodes to quantify the trust level. In practice, these scores are useful for managing the risks in communicating with other nodes in a dynamic and hostile IoT network. For instance, in a mobile crowdsourcing scenario, wherein each node individually collects data and contributes to the system, TRM is essential to provide a first layer safeguard to determine the trustworthiness of the collected data~\cite{putra2021d}.

In the following subsections, we describe the inherent building blocks of a TRM with its salient properties. We also discuss how blockchain would enhance TRM in the context of IoT applications.

\begin{figure*}
    \centering
    \includegraphics[width=0.96\textwidth]{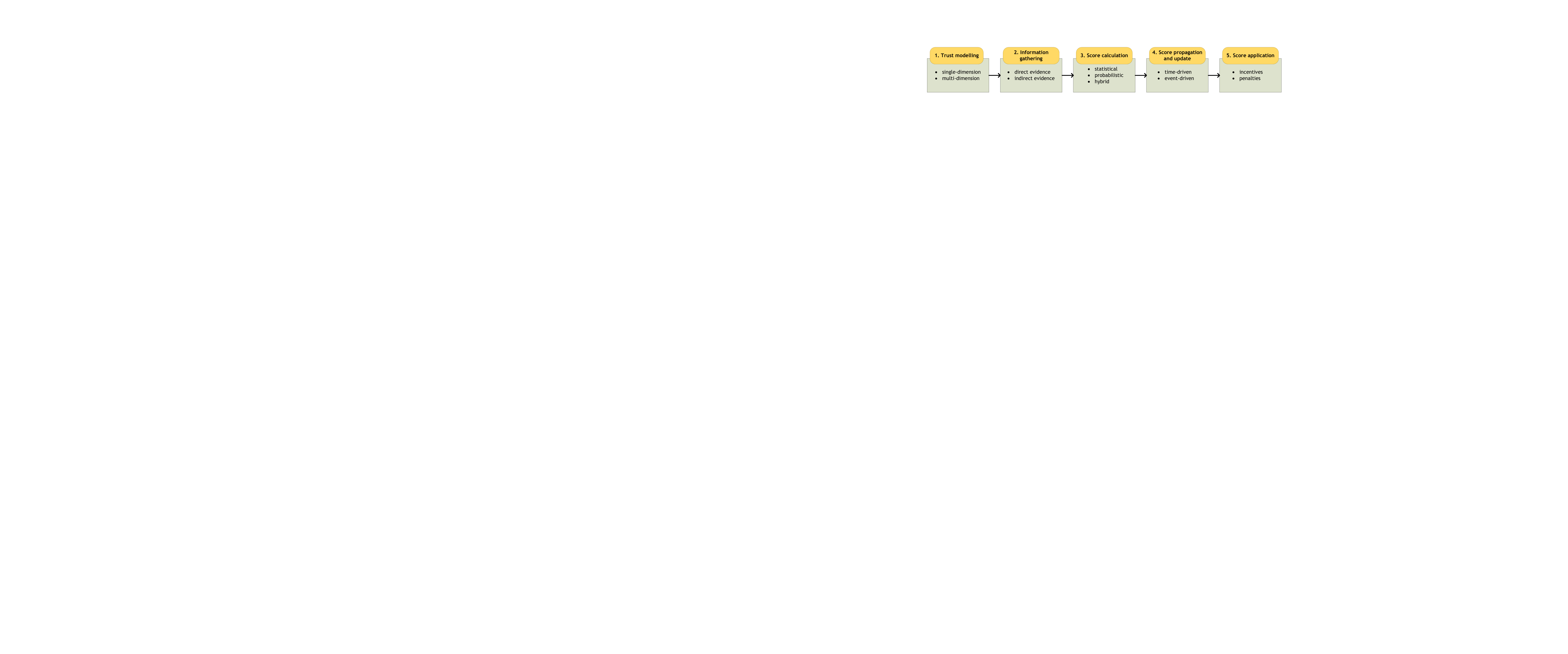}
    \caption{The five fundamental stages of TRM. While each IoT application has its own trust issues and unique challenges, the TRM follows these generic stages from modelling until the scores are applied to address the trust issues.}
    \label{fig:trm-evaluation-process}
\end{figure*}

\subsection{TRM Fundamentals}
\label{sub:fundamentals-trm}
In general, the evolution of trust and reputation follows similar pattern as in real world social interactions, wherein, these properties are gradually developed as more interactions happen. Naturally, these interactions may increase or decrease the trust level, depending on the perceived interaction experience. In the following, we describe the fundamental properties of trust and also the stages in TRM from trust modelling to the application of trust scores.

    \subsubsection{Trust Dimensions}
    \label{subsub:dimension}
    Depending on specific applications, TRM can model the trust using a simple single approach where trust is derived from single attribute. TRM can also use multi-dimensional approach where more detailed considerations are taken into account~\cite{Bellini2020}.

    \begin{itemize}
        \item \textbf{Single-dimension.} Single-dimension trust model could be an option where there is a need to focus on a specific metric to build trust. In addition, this type of trust model may be the best option if the network consists of mainly resource-constrained nodes.
    
        \item \textbf{Multi-dimension.} To capture multiple aspects of trust, multi-dimension trust model can be applied, where trust is derived from multiple attributes with certain weightings to put emphasise on several attributes over another. In practice, this model is preferred as it provides a richer representation of trust.
    \end{itemize}

    \subsubsection{Types of Trust}
    \label{subsub:trust-type}
    In general, IoT applications rely on the collection of sensor observations and the interactions between nodes. As such, trust can be generally classified into three main categories as follows.
    
    \begin{itemize}
        \item \textbf{Data-based.} In this category, trust is derived and calculated from the quality of the observation data, which is prone to noise, sensor drift, or bias. Here, the trust is directly correlated with the veracity and the authenticity of the data. Therefore, data-based trust relies heavily on data validation, which may include data correlation with other nearby sensor observations to validate if there is any degradation of the data quality.
    
        \item \textbf{Behaviour-based.} On the other hand, behaviour-based trust puts more emphasis on trusting the behaviour of the nodes. For instance, positive interaction is defined as a behaviour that conforms to the expected behaviour, which would increase the trust level. Conversely, a negative experience would degrade the trust level. Here, policy enforcement plays a significant role as a means to validate node's behaviour.

        \item \textbf{Hybrid.} In certain scenarios, utilising one trust category may not be adequate to derive trustworthiness. For example, supply chain applications necessitate that trusted traders are seen as the parties who supply reliable data and conform to the pre-agreed trade rules. In this scenario, trust and reputation could be evaluated with a consideration of both the data-based and behaviour-based approach, where weightings can be used to balance these two trust approaches.
    \end{itemize}

    \subsubsection{Fundamental Stages in TRM}
    \label{subsub:stages-TRM}
    TRM empirically evaluates trust in five sequential stages, shown in Fig.~\ref{fig:trm-evaluation-process}. We discuss the stages in more details as follows.

    \begin{enumerate}
        \item \textbf{Trust modelling.} The initial stage in TRM deals with how the trust and reputation should be quantified, which defines the attributes and parameters for trust quantification. The trust model oftentimes is application specific and cannot be generalised. Some attributes include the veracity of the sensor observation correlated with adjacent sensor data and the adherence of a node to the pre-determined policies.
        
        \item \textbf{Information gathering.} Subsequently, TRM gathers necessary information as per the trust model, using which the trust and reputation for each node are later evaluated. In general, TRM collects the information from two sources: 1) direct experience with a particular node; and 2) in the absence of prior interaction, TRM collects recommendation or feedback from other nodes.
        
        \item \textbf{Score calculation.} TRM could utilise multiple techniques to calculate the trust and reputation scores, depending on the pre-defined trust model and collected trust information. In blockchain-enabled TRM, smart contracts may be used to calculate the trust and reputation score in a transparent manner, which we explain in more detail in Section~\ref{sub:blockchain-properties}.
        
        \item \textbf{Score propagation and update.} The trust and reputation scores should be propagated to all participating nodes in the network, such that everyone is notified of the new scores. In addition, TRM may follow 1) time-driven update, in which the scores are continuously updated on regular basis, or 2) event-driven approach, where the scores are updated on certain events. The propagation may follow a centralised or distributed model, depending on the application architecture. In blockchain-based TRM, the score propagation occurs in parallel with block propagation, as the updated scores are stored in each new block.
        
        \item \textbf{Score application.} Finally, the trust and reputation scores could be applied to certain IoT scenario. For instance, the scores can be used as the basis to apply incentives and penalty mechanisms. We discuss various applications of TRM in more detail, based on our experience, in Section~\ref{sec:applications}.
    \end{enumerate}

\subsection{Blockchain Properties for TRM}
\label{sub:blockchain-properties}
As an integral component in blockchain-enabled IoT, TRM can benefit from the incorporation of blockchain, such as the transparent and deterministic nature of smart contracts, which we discuss in the following.

\begin{itemize}
    \item \textbf{Decentralisation.} In blockchain-enabled TRM, there is no centralised authority that manages the collection of trust evidence and evaluation of trust scores. That way, the risks of having a single point of failure or malicious authority can be effectively eliminated. Blockchain is managed in decentralised manner, wherein all nodes are participating in the process, with consensus algorithm as the safeguard to any fraudulent actions. Blockchain introduces a distributed ledger where the trust data is stored in an immutable and transparent fashion.
    
    \item \textbf{Immutability.} The inherent data structure of the distributed ledger, which consists of inter-linked blocks with hash cryptography, makes it difficult if not almost impractical to tamper with the data. TRM can benefit for this immutability property as the trust evidence could be securely stored without any risk of data alteration.
    
    \item \textbf{Transparency.} The transparent nature of blockchain would allow for reliable auditability, as the distributed ledger could act as a traceable source for auditing the trust calculation with the corresponding trust evidence. In addition, the smart contract provides a transparent trust calculation.
    
    \item \textbf{Smart contract.} Smart contracts allow for a deterministic and trusted execution of business logic in a decentralised manner. In blockchain-enabled TRM, smart contracts play a significant role of enforcing the TRM business logic in a transparent manner. For example, an IoT node may submit feedback about its recent experience with a service provider, after which the smart contract updates the provider's trust and reputation score.
    
    \item \textbf{Pseudonymity.} Blockchain utilises public-key cryptography instead of real identities for its authentication mechanism. An IoT node is thus recognisable by its public key, which to some extent protects its privacy, as the real identities are not exposed to public.
\end{itemize}

\section{Applications of TRM for IoT}
\label{sec:applications}
In this section, we discuss our research experiences in TRM for blockchain-enabled IoT. We first discuss specific trust issues in different application domains and present TRM frameworks to address them. We then identify the unique challenges of incorporating TRM to these application domains, which influence the design of the TRM framework. We present a summary of our experiences in Fig.~\ref{fig:journey-summary}.

\begin{figure}
    \centering
    \includegraphics[width=0.49\textwidth]{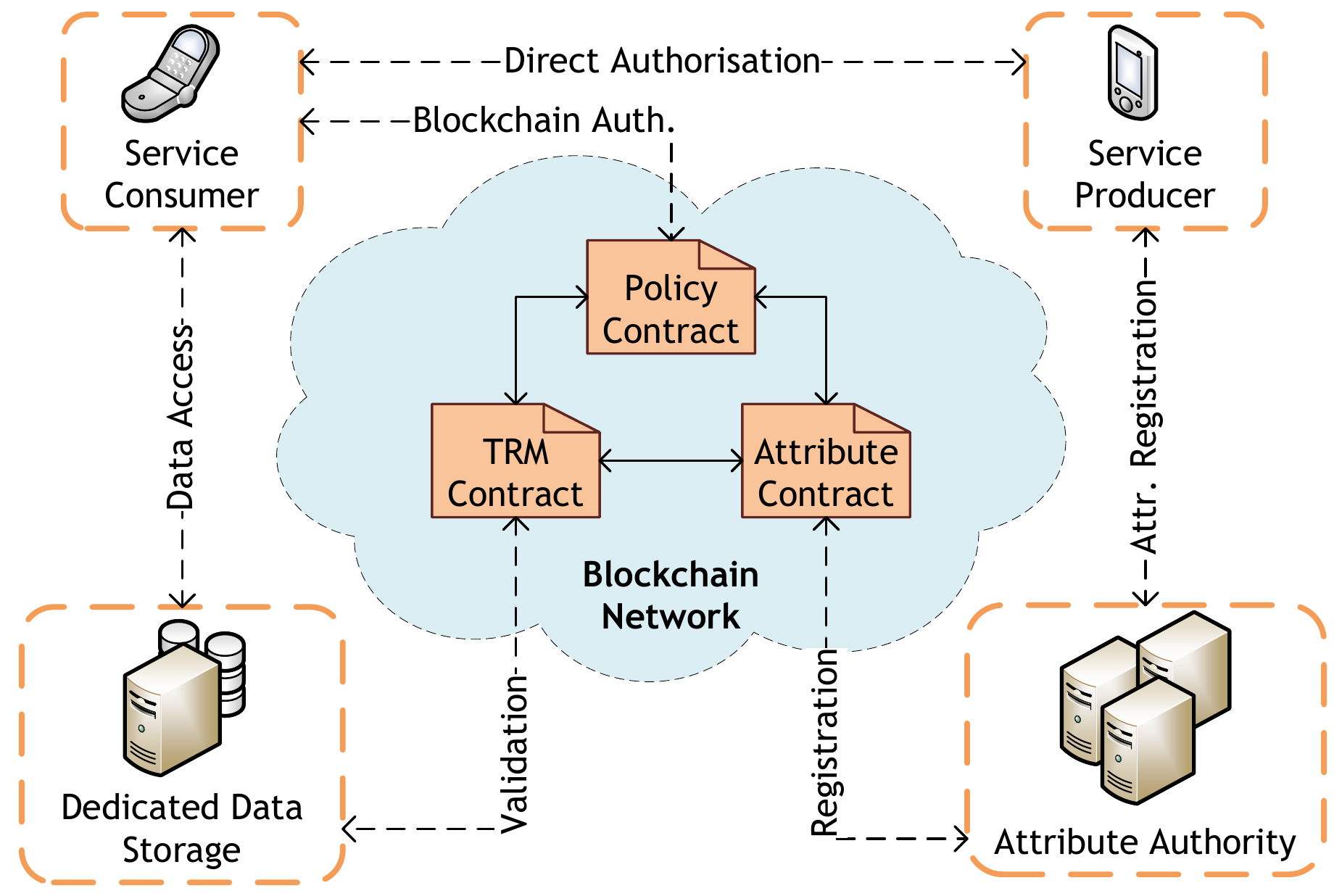}
    \caption{Blockchain-based TRM for dynamic access control schemes in IoT~\cite{putra2020}. Blockchain is an instrumental building block where smart contract is utilised to enforce transparent and auditable trust calculation and access control enforcements.}
    \label{fig:trm-access-control}
\end{figure}

\begin{figure}
    \centering
    \includegraphics[width=0.49\textwidth]{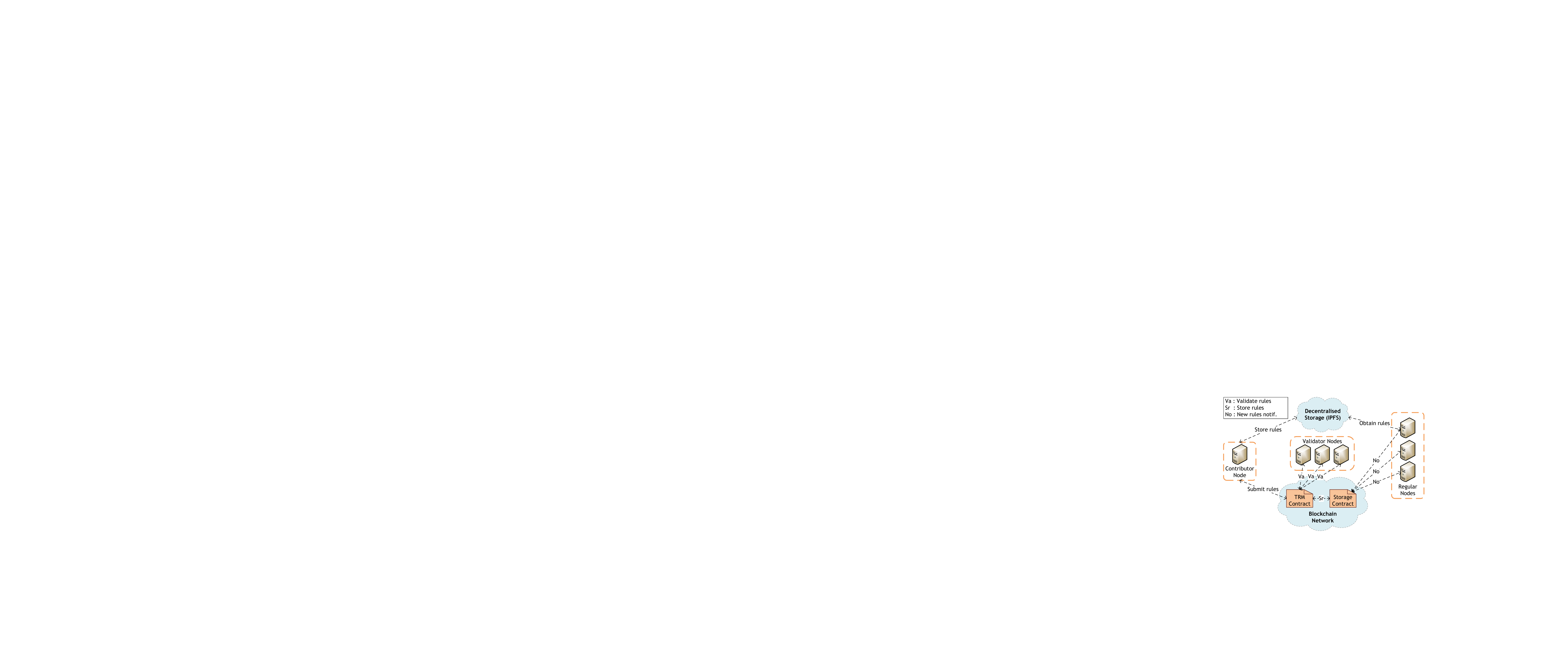}
    \caption{The design of our trustworthy CIDS framework~\cite{putra2021c}, where TRM plays an important role in validating contributed detection rules and assigning score to each of the validated rules.}
    \label{fig:cids-arch}
\end{figure}

\subsection{IoT Access Control}
\label{sub:access-control}
Access control aims to enforce restrictions on the actions that authenticated users can execute on a particular IoT system~\cite{sandhu1994access}. With access control, or authorisation, any illegitimate access to protected resources or security breaches can be avoided. For instance, resource owners can define an access policy to a particular resources in IoT networks, wherein other nodes need to authorise themselves to access the resource.

\textbf{Challenges.} There have been several proposals for blockchain-based IoT access control schemes~\cite{Jiang2018, Pinno2018}, where the schemes work by enforcing static access policies, based on an assumption that the authenticated nodes would always show benign behaviour. However, authenticated nodes may be compromised post-authentication and become malicious, which may significantly disrupt the network. A prominent example is the extensive damage caused by the Mirai botnet~\cite{antonakakis2017}. This motivates the need to design dynamic access control mechanisms, without overlooking the fact that authorisation demands strict consideration of certain attributes to access a resource.

\textbf{Contributions.} To address the aforementioned challenges, we proposed a trust-based blockchain authorisation for IoT~\cite{putra2020} (see Fig.~\ref{fig:trm-access-control}), where we utilise TRM to supply additional attributes to a decentralised Attribute-Based Access Control (ABAC) mechanism.
Our TRM framework progressively quantifies the trustworthiness of both Service Consumers (SC) and Providers (SP) into trust and reputation scores, using which we define more attributes to enrich the ABAC mechanism, hence achieving dynamic access policies.
In general, the trust and reputation scores are calculated based on the nodes' adherence to the access control policies. In our definition, trust is a subjective belief on SC's past behaviour, which may aid in forecasting SC's future behaviour. On the other hand, reputation is seen as the public view of SC's past behaviour from multiple trust sources. We define three smart contract to manage the access control mechanism in a transparent and auditable manner, namely policy, attribute and TRM contracts, as seen in Fig~\ref{fig:trm-access-control}. Policy contract stores and enforces the access policies based on the required attributes, while attribute contract is the point of contact when policy contract needs to verify some attributes; and TRM contract calculates and updates the trust and reputation scores. In our framework, we note that continuous evaluation of the nodes behaviour can be a potential way to detect and eliminate compromised nodes in the IoT network~\cite{MohamadNoor2019}. Our solution is implemented on a private Ethereum network, using which we benchmark our TRM framework. The experimental results, which include the latency, score evolution and gas consumption of the smart contracts, indicate the applicability of our solution in the IoT context.

In~\cite{putra2021b}, we proposed an improvement of our dynamic access control framework, where we emphasise privacy preservation and efficient trust computation. To conceal users' personal information from public access, where we utilise multiple private side-chains to privately store sensitive users' attributes. We conceive a lightweight TRM model that reduces the trust computation load by using a simple recursion. Our improved framework is implemented on a public Rinkeby Ethereum test-network interconnected with a lab-scale testbed of Raspberry Pi computers. The evaluations consider various performance metrics to highlight the applicability of the proposed solution for IoT contexts.

\begin{figure}
    \centering
    \includegraphics[width=0.46\textwidth]{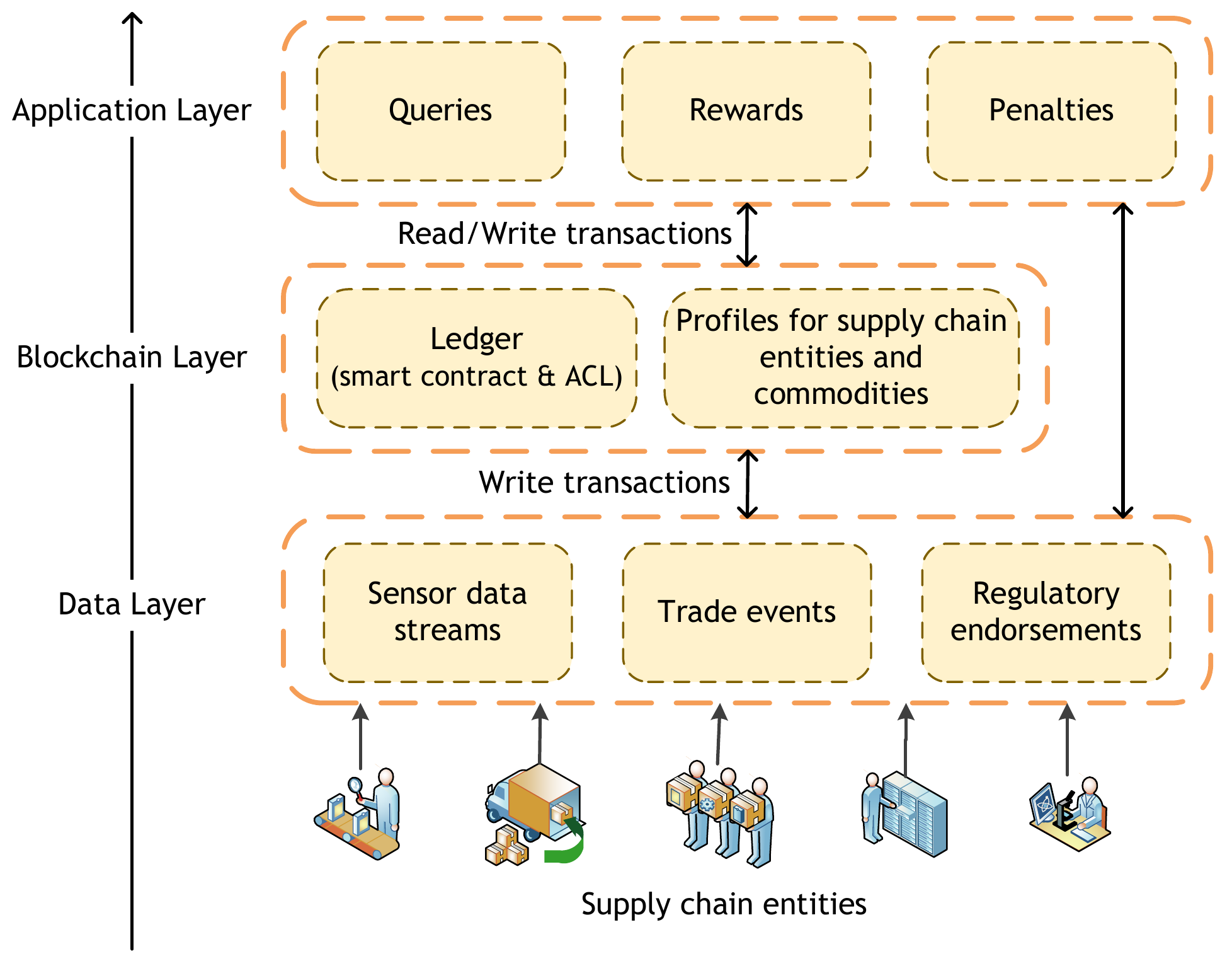}
    \caption{The layered architecture of TrustChain~\cite{malik2019}, which constitutes three interconnected layers. Data layer supplies multi-faceted observation from which the trust and reputation scores are derived.}
    \label{fig:trustchain-arch}
\end{figure}

\begin{figure}
    \centering
    \includegraphics[width=0.46\textwidth]{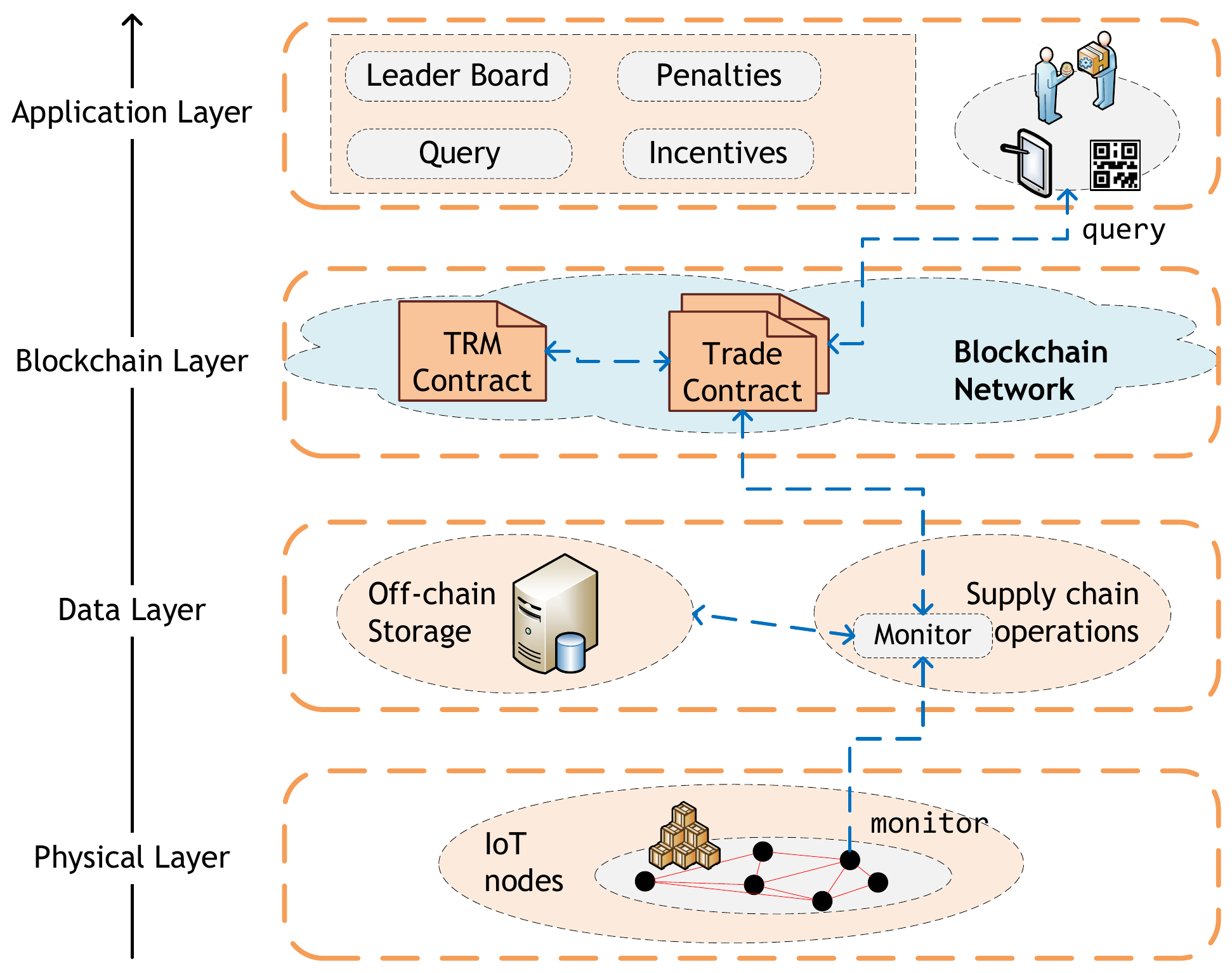}
    \caption{An architecture overview of DeTRM~\cite{dharmaputra2022}, which consists of four layers. The modular design of DeTRM allows for more applications as other application can be built on top of the blockchain layer.}
    \label{fig:detrm-arch}
\end{figure}

\subsection{Intrusion Detection System}
\label{sub:ids}
Intrusion Detection Systems (IDS) have been widely used in the industry to alert the network administrator of any possible threats in the IoT networks~\cite{putra2021c}.
The proliferation of IoT devices in diverse application domain has consequently enlarge the attack surface, which has raised the urgency to defend the network against emerging attacks~\cite{Anthi2019}.
In recent years, blockchain has prompted the new notion of decentralised Collaborative IDS (CIDS), wherein blockchain enables IDS nodes to collaboratively contribute relevant information, such as intrusion detection rules, which helps increase the overall capability of the IDS nodes to detect unknown threats~\cite{Li2019}.

\textbf{Challenges.} However, the existing work of blockchain-based CIDS generally presume that the contributor nodes are continually honest and overlooks the necessity to regularly evaluate the trustworthiness of the participating IDS nodes. Without proper trust evaluation in CIDS, compromised nodes may share deceptive detection rules on purpose, which would jeopardise the network to detrimental attacks~\cite{Kolokotronis2019}.

\textbf{Contributions.} To address the trustworthiness issues in CIDS, we proposed a trustworthy blockchain-based CIDS framework~\cite{putra2021c}. In our proposed framework, the TRM progressively evaluates the trust level of each CIDS node by assessing the quality of the contributed detection rules. In Fig.~\ref{fig:cids-arch}, we group the CIDS nodes into three categories, namely contributor, validator and regular nodes. We introduce the notion of validator nodes, who perform off-chain assessment of the submitted detection rules. Subsequently, validator nodes assign numerical scores to each contributed rule, using which the trustworthiness level of CIDS node is determined. The TRM framework employs smart contracts to transparently update and calculate the scores. As such, each CIDS node can conveniently update its local detection database with the validated detection rules. To achieve decentralisation in the storage of detection rules, we utilise a peer-to-peer storage system, such as The InterPlanetary File System (IPFS)~\cite{benet2014ipfs}. Our framework is designed to be blockchain-agnostic, such that it can be implemented on any blockchain platforms with support in smart contracts. 
Signature-based CIDS is used as an illustrative example,
but our TRM framework can be also generalised to other CIDS types, such as Machine Learning (ML) or anomaly-based. We tested our proposed framework on a lab-scale Ethereum private network, where results indicate that our proof-of-concept 
performs within the expected benchmarks of Ethereum, highlighting its feasibility.

\subsection{Supply Chain}
\label{sub:supply-chain}
Supply Chain Management Systems (SCMS) provide an end-to-end tracking of the traded commodities, which spans from sourcing the raw materials to the final retail shops~\cite{juma2019}. SCMS create digital records of the traded goods, referred to as digital assets, using which SCMS tracks the transfers of commodities and ownership changes. IoT has been implemented to improve SCMS~\cite{rejeb2019} by information gathering at various stages of supply chain, for instance sensor networks that provide automated monitoring of commodity storage. In recent years, blockchain has shown its potential to enhance conventional SCMS due to its inherent data structure of immutable time-stamped records~\cite{malik2019}. Supply chain activities such as sourcing goods, trading, manufacturing, ownership transfer etc., can thus be logged on blockchain in an immutable way. This increases transparency of products as they move through their life cycle. In addition, blockchain's smart contracts are able to automate certain operations, such as, provenance, compliance checking and trust computation, based on supply chain events and activities in a secure and verifiable manner. 

\textbf{Challenges.} However, blockchain alone cannot ascertain the authenticity and veracity of the IoT observations provided by supply chain participants. In fact, IoT nodes may turn faulty from time to time and thus record and send inaccurate observation. As such, the authenticity and the integrity of the supply chain data become questionable. The fact that the recorded data becomes immutable once stored on blockchain exacerbates this trust issue on supply chain data, as the ledger would be contaminated with bad data~\cite{powell2021a}.

\textbf{Contributions.} To address the trust issues in blockchain-enabled supply chain, we argue that TRM would be an effective solution. In~\cite{malik2019}, we proposed TrustChain, where we derive and evaluate trust from multiple sources, namely sensor observations, regulatory inspections and trade events, as seen in Fig.~\ref{fig:trustchain-arch}. TrustChain provides flexibility in calculating the trust scores at different perspective, for instance, at the commodity level or a particular supply chain participant to evaluate its role in the supply chain. Thus we assign trust scores to the traded commodities and the traders both. We use two types of smart contracts, namely quality and rating contracts, to enlist certain quality criteria for the corresponding commodity, e.g., temperature boundary and contract identifier, and to compute the scores after a trade event, respectively. The use of smart contracts enables transparent automation of trust score calculation, using which we apply reward and penalty mechanisms to encourage truthful contribution of data to the blockchain. Our proof-of-concept implementation of TrustChain in Hyperledger Fabric reveals that our proposed trust mechanisms for supply chain only incur minimal overheads in throughput and latency.

While TrustChain was among the first proposals in incorporating TRM for IoT-enabled supply chains, we note that TrustChain assumes that there are no alterations of commodities as they make their way from the primary producers to the retail shelves. However, this assumption is unlikely in real world supply chain scenarios, as it is quite common to have production processes somewhere in the supply chain, such as production of cheese products from raw milk. In addition, TrustChain fully assigns blockchain operation and maintenance to a business network administrator, which makes the use of blockchain questionable as it is essentially a centralised TTP.

To address the issues, we proposed DeTRM~\cite{dharmaputra2022}, where we design a tailored TRM framework for SCMS by considering inherent supply chain operations into our trust model, for instance, production of new products from existing commodities in the SCMS. Therefore, DeTRM tracks all trust information within the entire supply chain life cycle, i.e., from sourcing the raw material to the retail shelf. To derive trustworthiness, DeTRM correlates sensor observations to evaluate the data quality into numerical measures and assesses the behaviour of supply chain entities against the pre-agreed trade contracts. To establish a solid decentralised framework, we require all supply chain entities, such as producers, traders and distributors, to run blockchain peer nodes for maintaining a network of consortium blockchain. In our framework, we exclude the end product customers from the blockchain network, but they could track the trust and provenance information of the products by scanning the imprinted QR code, see Fig.~\ref{fig:detrm-arch}. The proof-of-concept implementation of DeTRM in a lab-scale Hyperledger Fabric network highlights that our framework incurs minimal overheads, with regards to resource utilisation, latency and throughput.

\begin{figure}
    \centering
    \includegraphics[width=0.47\textwidth]{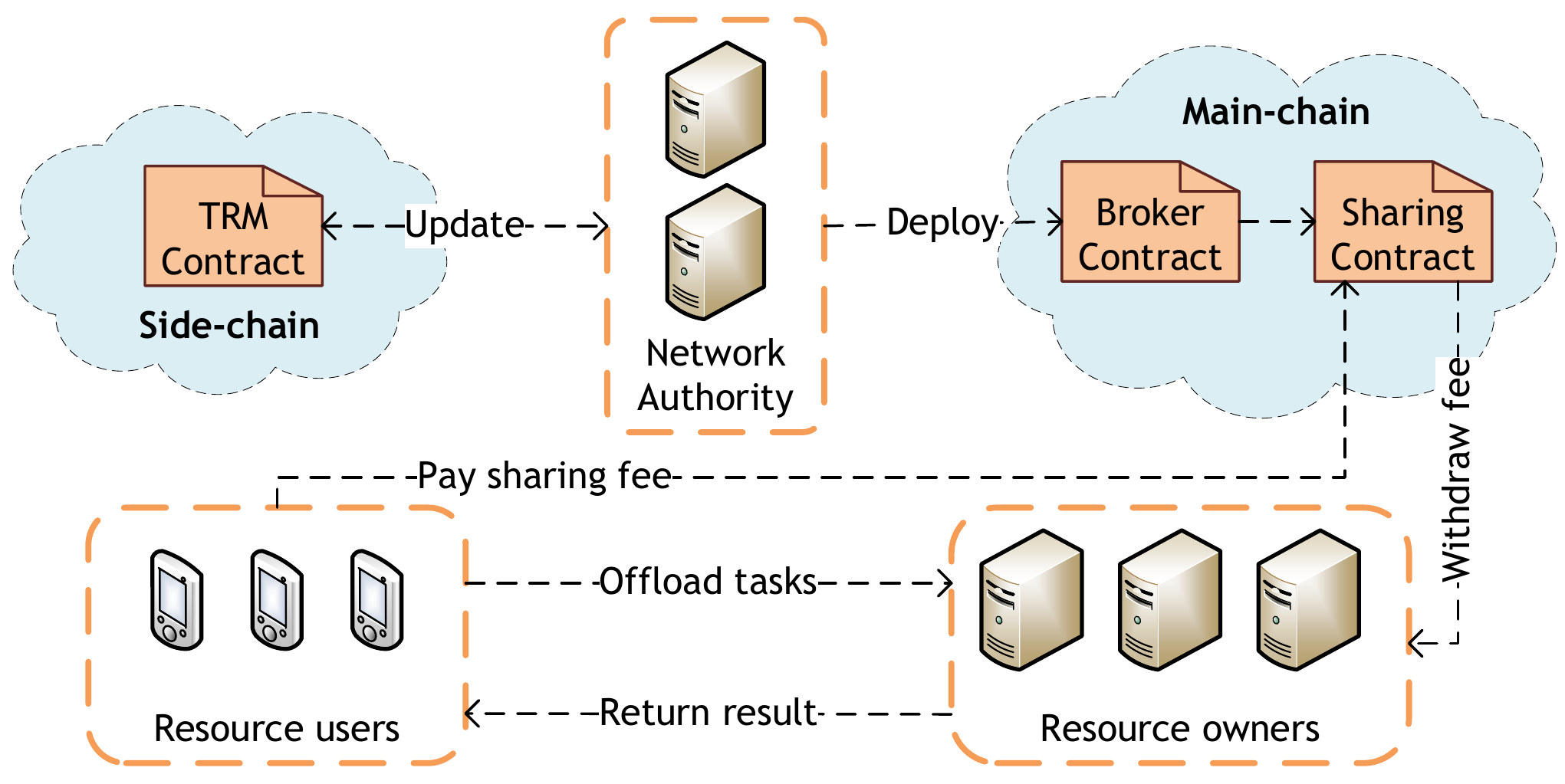}
    \caption{The overview of our resource sharing mechanism~\cite{putra2022a}, where the resource users offload computation tasks to resource owners. Here, the side-chain hosts the TRM contract, while the main-chain records the sharing agreements via broker and sharing contracts.}
    \label{fig:res-sharing-arch}
\end{figure}

\subsection{IoT Resource Sharing}
\label{sub:6g-iot}
Resource sharing in IoT network is intended to achieve effective utilisation of scarce network resources, such as radio spectrum and computation infrastructures~\cite{wang2021}. Resource sharing can also alleviate the computation workload on constrained IoT devices, such as edge computing~\cite{pang2021}. Blockchain has received significant attention from research and industry communities as an integral building block to realise the Dynamic Resource Sharing (DRS)~\cite{hu2021}. Due to the verifiable automation enabled by blockchain, DRS can significantly improve the resource utilisation, compared to the conventional resource allocation schemes, which tend to be static and inflexible. Blockchain can effectively act as a trustless intermediary to facilitate communications between parties in resource sharing, making the coordination and cooperation more effective and efficient.

\textbf{Challenges.} However, blockchain-based resource sharing for IoT still suffers from several challenges. While blockchain provides consistency and trust in the stored data within a resource sharing environment, blockchain alone cannot ascertain the trust in the participating nodes when performing resource sharing. The lack of trust would discourage nodes in the network to share or use other resources, as some malicious nodes may present and can potentially launch adverse attacks. These trust issues would undermine the initial goal of resource sharing, as its effectiveness is reduced. 

\textbf{Contributions.} To address the above mentioned trust issues, we proposed a TRM framework to support resource sharing, which aims to encourage cooperation in resource sharing by evaluating participants' reliability after each sharing instance~\cite{putra2022a}. We consider the use case of computation resource sharing in edge computing, where the resource users offload computation tasks to resource owners (see Fig.\ref{fig:res-sharing-arch}), after which the trustworthiness of the resource owners are determined. Our framework also allows for both resource owners and users to employ changeable keys to obfuscate their transaction traces in the network, avoiding the de-anonymisation attacks that may reveal sensitive information of critical network infrastructures. Our framework exploits smart contracts to provide auditable trust calculation, where a recursive trust computation is proposed to meet the efficiency requirements of trust calculation in 6G networks. Our framework demonstrates how blockchain, with the aid of a TRM, could provide reliable assurance in the trust between participating nodes in the network, thus reducing the inherent risk of resource sharing. We developed a proof-of-concept implementation of our proposed framework on lab-scale private networks. The experimental results signify the feasibility of our framework, as it only incurs minimal overheads with regards to gas consumption and overall latency.

\section{Discussion}
\label{sec:discussion}
In this section, we summarise the lessons learned from our journey in incorporating blockchain-based TRM in various IoT applications. We also discuss several future research directions for TRM in blockchain-enabled IoT.

\subsection{Lessons Learned}
\label{sub:lessons-learned}
We note that there are four lessons from our research in blockchain-based TRM, which are discussed in the following.

\begin{itemize}
    \item \textbf{Trust model.} Various IoT applications, discussed in Section~\ref{sec:applications}, point out that each application requires a specific trust model that is influenced by its unique challenges. While one can borrow a trust model conception from a particular domain, the model still needs to be adjusted to match with the way trust is evaluated in the adopted domain.
    \item \textbf{Smart contracts.} With the incorporation of blockchain for TRM, smart contracts become an instrumental building block, where the main TRM logic is implemented. All IoT applications in Section~\ref{sec:applications} utilise smart contracts to transparently collect trust evidence and calculate the corresponding trust and reputation scores. Smart contract also allows for better auditability in the TRM process, which bring more confidence in the TRM framework as all TRM operations can be transparently tracked on-chain.
    \item \textbf{Off-chain component.} While blockchain-based TRM benefits from transparent on-chain computation, some underlying TRM processes cannot be implemented on-chain due to their computation complexity. In certain conditions, off-chain computation is required to complement the smart contract, as demonstrated in our TRM framework for CIDS. In addition, blockchain may not be an ideal storage medium for large amount of data, as in the case of supply chains, which highlights the need to incorporate off-chain storage, such as cloud or peer-to-peer storage.
    \item \textbf{Changeable keys.} Blockchain-based TRM benefits from the utilisation of pseudonyms, which conceal the node's real identity. However, it possible to de-anonymise the pseudonyms~\cite{kang2020}, especially if the public keys are used repeatedly. IoT nodes may opt to use changeable keys in different transactions. However, the TRM needs to be aware that multiple keys correspond to the same node, when calculating the trust and reputation scores.
\end{itemize}

\subsection{Future Research Directions}
\label{sub:future-work}
While there have been several works that have developed TRM for blockchain-enabled IoT, we note that there are still several open challenges, which include security, privacy, interoperability and scalability.

\textbf{Security.}
It is known that TRM is susceptible to attacks, wherein the adversaries attempt to illegitimately increase their reputation score and ruin others. Incorporation of blockchain to TRM may render the conventional defence schemes unusable, which highlights the need to design a robust blockchain-based TRM against these trust-related attacks. Some common attacks include: 1) \textit{Sybil attack}, where adversaries create fake identities to increase or ruin trust scores to their benefit; 2) \textit{ballot-stuffing}, where a node illegitimately increases its score, for instance by colluding with other nodes; and 3) \textit{bad-mouthing}, where the adversaries attempt to ruin honest users' reputation scores, for instance by submitting fake negative feedback. Further research should also study the possible impact on the network reliability if these security measures fail.

\textbf{Privacy.} As discussed in Section~\ref{sub:lessons-learned}, there is the need to design privacy-preserving TRM using changeable keys. Zero-Knowledge Proof (ZKP) is a potential technique for concealing real identities, which works by validating a given mathematical proof without revealing any information that could be traced back to the user's real identity. In ML-based TRM~\cite{putra2022a}, federated learning could be an approach to achieve privacy-preservation as some trust related data may reveal sensitive information. In addition, future research also needs to study the mechanisms to conceal feedback information, which would encourage IoT nodes to submit truthful feedback.

\textbf{Interoperability.} Future IoT networks, for instance 6G-enabled IoT, will offer integration of autonomous networks, which may run an independent TRM~\cite{putra2022a}. To allow collaboration between these autonomous networks with their own TRM, future work should aim to devise a reliable and trustworthy trust and reputation scores transfer between independent TRMs. In addition, TRM often experience the lack of prior interactions which may hinder the calculation of trust and reputation scores. With interoperable TRMs, the absence of prior interactions can be addressed by transferring trust or scores from another TRM that has already collected trust evidence about a particular entity. To this end, future research should also aim to provide cross-chain data transfer for transferring trust evidence between independent blockchains.

\textbf{Scalability.} Blockchain-based TRM inherently suffers from limited blockchain throughput, which makes it difficult to scale. As the throughput issues actually stem from the resource consumptive consensus algorithm, future research should aim to provide fast consensus algorithm, such as Proof-of-Stake, which has recently become the default consensus algorithm for Ethereum~\cite{ethereumpos2022}. The use of fast and efficient consensus can help TRM scale, especially in large-scale IoT deployments. In addition, future research is required to enhance the scalability of the append-only shared ledger, which can grow exponentially as old trust and reputation data would remain in the ledger as the data for new interactions continues to be recorded on-chain.

\section{Conclusion}
\label{sec:conclusion}
In this paper, we presented our research journey over the past few years, where we address trust issues in various IoT applications, including access control, intrusion detection systems, resource sharing and supply chains. We noted that each application has its own trust issues and unique challenges which influence the design of the corresponding TRM. We also discussed the lessons we learned in the process and highlighted several open research challenges in blockchain-based TRM research for future work, which include security, privacy, interoperability and scalability.

\section*{Acknowledgement}
The work has been partially supported by the Cyber Security Research Centre Limited whose activities are partially funded by the Australian Government’s Cooperative Research Centres Programme.

\bibliographystyle{IEEEtran}
\bibliography{./ref.bib}

\end{document}